\documentclass[twocolumn,amsmath,amssymb,prapplied]{revtex4-2}
\usepackage{latexsym}
\usepackage{graphicx}
\usepackage{psfig}
\usepackage{color}
\usepackage{verbatim}
\usepackage{multirow}
\usepackage[letterpaper, margin=1in]{geometry}
\usepackage{siunitx}
\usepackage[capitalise]{cleveref}
\usepackage{circledsteps}
\usepackage{appendix}

\begin{document}

\preprint{}
\title{Characterizing losses in InAs two-dimensional electron gas-based gatemon qubits}
\author{William M. Strickland$^{1}$}
\author{Lukas J. Baker$^{1}$} 
\author{Jaewoo Lee$^{1}$}
\author{Krishna Dindial$^{1}$} 
\author{Bassel Heiba Elfeky$^{1}$} 
\author{Patrick J. Strohbeen$^{1}$}
\author{Mehdi Hatefipour$^{1}$} 
\author{Peng Yu$^{1}$} 
\author{Ido Levy$^{1}$}
\author{Jacob Issokson$^{1}$}
\author{Vladimir E. Manucharyan$^{2}$}
\author{Javad Shabani$^{1}$}

\affiliation{$^{1}$Center for Quantum Information Physics, Department of Physics, New York University, New York 10003, USA}
\affiliation{$^{2}$Institute of Physics, École Polytechnique Fédérale de Lausanne, 1015 Lausanne, Switzerland}
\date{\today}

\begin{abstract}
The tunnelling of cooper pairs across a Josephson junction (JJ) allow for the nonlinear inductance necessary to construct superconducting qubits, amplifiers, and various other quantum circuits. An alternative approach using hybrid superconductor-semiconductor JJs can enable superconducting qubit architectures with all electric control. Here we present continuous-wave and time-domain characterization of gatemon qubits and coplanar waveguide resonators based on an InAs two-dimensional electron gas. We show that the qubit undergoes a vacuum Rabi splitting with a readout cavity and we drive coherent Rabi oscillations between the qubit ground and first excited states. We measure qubit relaxation times to be $T_1 =$  100 ns over a 1.5 GHz tunable band. We detail the loss mechanisms present in these materials through a systematic study of the quality factors of coplanar waveguide resonators. While various loss mechanisms are present in III-V gatemon circuits we detail future directions in enhancing the relaxation times of qubit devices on this platform. 

\end{abstract}

\pacs{}
\maketitle

\begin{figure}
    \centering
    \includegraphics[width=0.45\textwidth]{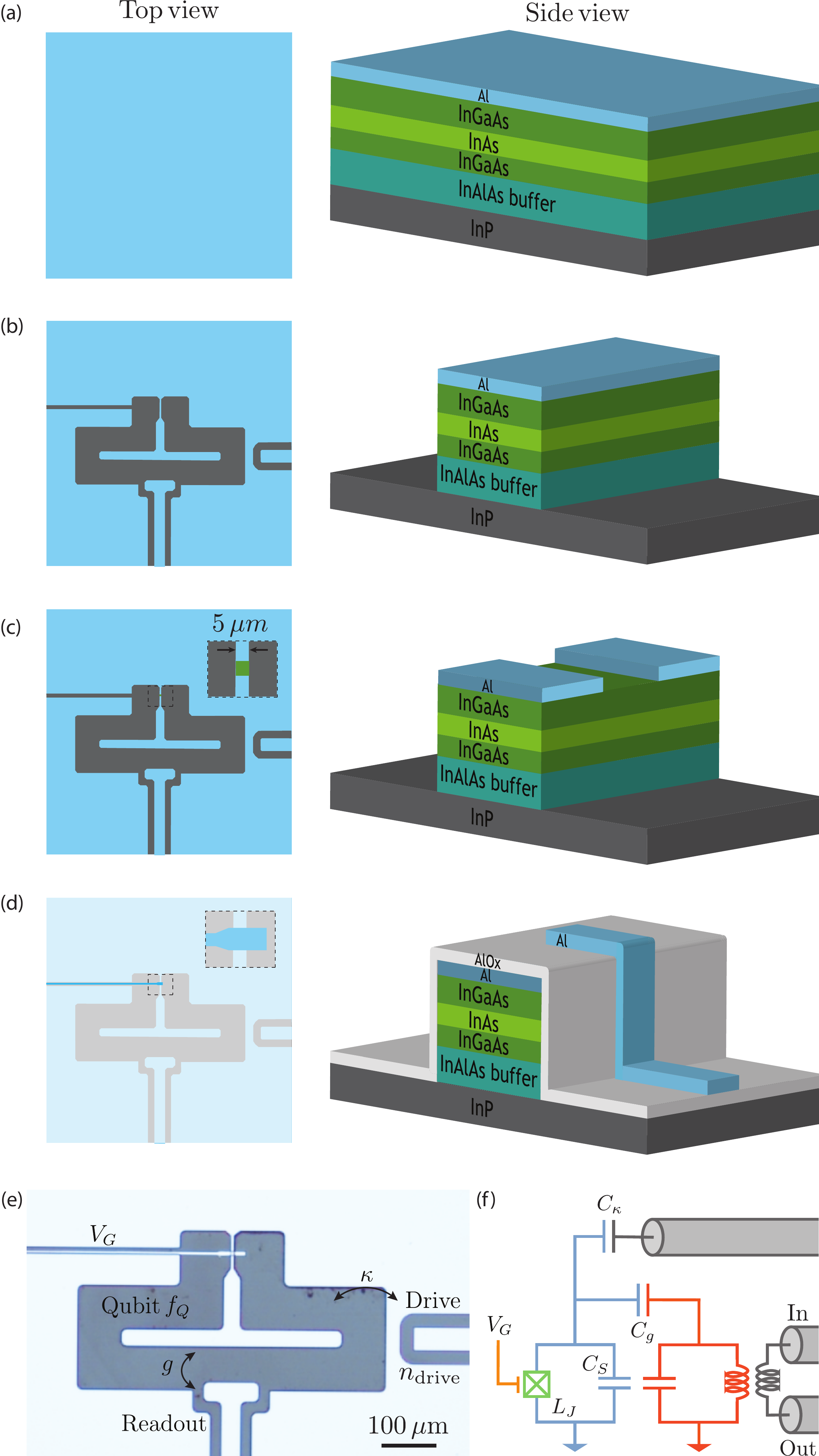}
    \caption{\textbf{Fabrication procedure and optical image:} The left column shows a top view of the surface, while the right column shows the side view, where each layer is clearly visible and noted. The layer structure is shown in (a) where layers of InAlAs (teal), InAs/InGaAs (green/dark green), and Al (blue) are grown on an InP substrate (gray).  The Al and III-V layers are etched in order to define the microwave circuit as shown in (b). The Josephson junction shown in (c) is then defined with an Al etch. The width of the superconducting electrodes are nominally 5 um and are separated by 100 nm (not to scale). We then blanket deposit a layer of AlO$_{x}$ (white), pattern the gate electrodes, and deposit an Al gate electrode, as shown in (d). An optical image of the device after fabrication is shown in (e). The qubit has a characteristic frequency $f_Q$ controlled by the gate voltage $V_G$ and is coupled to a readout resonator with a coupling strength $g$, set by the coupling capacitance $C_g$. An external drive line is coupled with a strength of $\kappa$, set by the capacitance $C_\kappa$, which drives transitions in the qubit. The equivalent circuit diagram is shown in (f). The Josephson inductance $L_J$ is shunted to ground by a capacitance $C_S$. Input and output lines are coupled inductively to the readout resonator (orange).}
    \label{fig:fab}
\end{figure}
\section{Introduction}
The superconducting qubit is a hallmark solid-state system that displays quantum coherence and strong light-matter coupling \cite{kjaergaard_currentstateofplay, wallraff_strong_2004, Koch2007, blais2004}. Recently, the coherence times of planar transmon qubits have exceeded \SI{300}{\micro s} \cite{place2021} and further improvements are expected with improved materials and fabrication \cite{deleon2021}.  A common design choice is to introduce flux tunability of a qubit or coupler for fast, high-fidelity single-qubit control and two-qubit gates \cite{chen2014, arute2019quantum}, almost exclusively realized by flux-sensitive superconducting quantum interference devices (SQUIDs) \cite{orlando1999, chiorescu2003,  palacios-laloy_tunable_2008, naaman_-chip_2016}. An architecture based on flux-biased SQUIDs may lead to future complications, however. The heat load induced by milliampere-level currents flowing through resistive wires can impose substantial cooling requirements as the scale of superconducting qubit chips increases. Stray magnetic fields in higher density qubit arrays can also cause irremediable crosstalk. In addition, low-frequency $1/f$-type flux noise can limit the dephasing times of flux-tunable qubits, and its origin and mitigation is an active area of research \cite{kakuyanagi2007, hutchings_tunable_2017, kumar2016, bialczak2007, rower2023}. An all-electric tunability scheme may prove to be beneficial in large-scale quantum processors, and JJs based on hybrid superconductor-semiconductor (S-Sm) materials are one interesting candidate to realize this.

A hybrid S-Sm Josephson junction device has current flow facilitated by Andreev bound states in the semiconductor weak-link. By biasing with an applied gate voltage one can tune the Fermi level in the semiconductor and the occupation of Andreev bound states, effectively controlling the conduction through the junction. It was shown that InAs makes an excellent candidate for the semiconductor in a hybrid S-Sm junction because it makes an Ohmic contact with superconducting metals such as Al \cite{mead1964}. It was discovered later that thin films of Al (111) can grow epitaxially on InAs (100) by molecular beam epitaxy \cite{Shabani2016, sarney_2020_metallization, sarney2018}, enabling a high quality contact \cite{Henri17, mayer2019, kjaergaard2017} on a wafer-scale. The incorporation of an S-Sm junction in a superconducting qubit was first demonstrated in Refs. \citenum{Larsen_PRL} and \citenum{deLange2015} using an InAs nanowire and since voltage tunable Josephson junctions have since made many appearances in qubits \cite{deLange2015, Larsen_PRL, Casparis2018, Casparis2016, kringhoj2018, wang_coherent_2019, kringhoj_parity2020, danilenko2022, yuan2021, Hays2020, hays2021, hertel2022}, couplers \cite{Maxim17, sardashti2020, hazard2022, chen2023voltage,  materise2023, casparis2019, splitthoff2022, strickland_superconducting_2023}, and other elements, such as an amplifiers and nonreciprocal elements \cite{phan2022, splitthoff2023gatetunable, leroux2022}. Of note, a wafer-scale architecture was implemented in the form of an InAs two-dimensional electron gases (2DEG) in Ref. \citenum{Casparis2018}, making the qubit processing more amenable to bottom-up fabrication. While, fabrication reproducibility and lossy III-V materials have prevented the wider adoption of the 2DEG-based gatemon qubit architecture, the prospect of using the voltage tunable junction in tunable couplers between qubits to implement low power, fast two-qubit gates could make near-term potentially useful and interesting material platform to study and integrate with state of the art superconducting qubits.

We report on the progress in improving the coherence properties of qubits based on hybrid S-Sm materials. The article is organized as follows: In Section II we report on coherent manipulation of InAs 2DEG-based gatemon qubits. The observed qubit frequency is tunable over \SI{1.5}{\giga Hz} and the qubit undergoes a vacuum Rabi splitting with the readout resonator. Coherent Rabi oscillations are driven between the ground and first excited states of the qubit and by fitting the decay of these oscillations we find that the characteristic time scale of the decay $T_2^\mathrm{Rabi}$ is 97 ns. The energy relaxation times $T_1$ of gatemon qubits over a wide gate voltage range are found with a maximum $T_1 = \SI{102}{\nano s}$, where $T_1$ generally increases with decreasing qubit frequency. In Section III, a systematic study of the various loss mechanisms in gatemon circuits, such as dielectric loss from two-level systems, and inductive loss, is performed by analyzing quality factors of coplanar waveguide (CPW) resonators. We measure a variety of samples with varying Al film thicknesses and find that the highest measured quality factor at low power is \SI{4e4}{}, suggesting that current qubit devices are limited by inductive loss in the superconductor. Finally in Section IV, we outline future steps to enhance gatemon $T_1$ times.

\begin{figure}
    \centering
    \includegraphics[width=.45\textwidth]{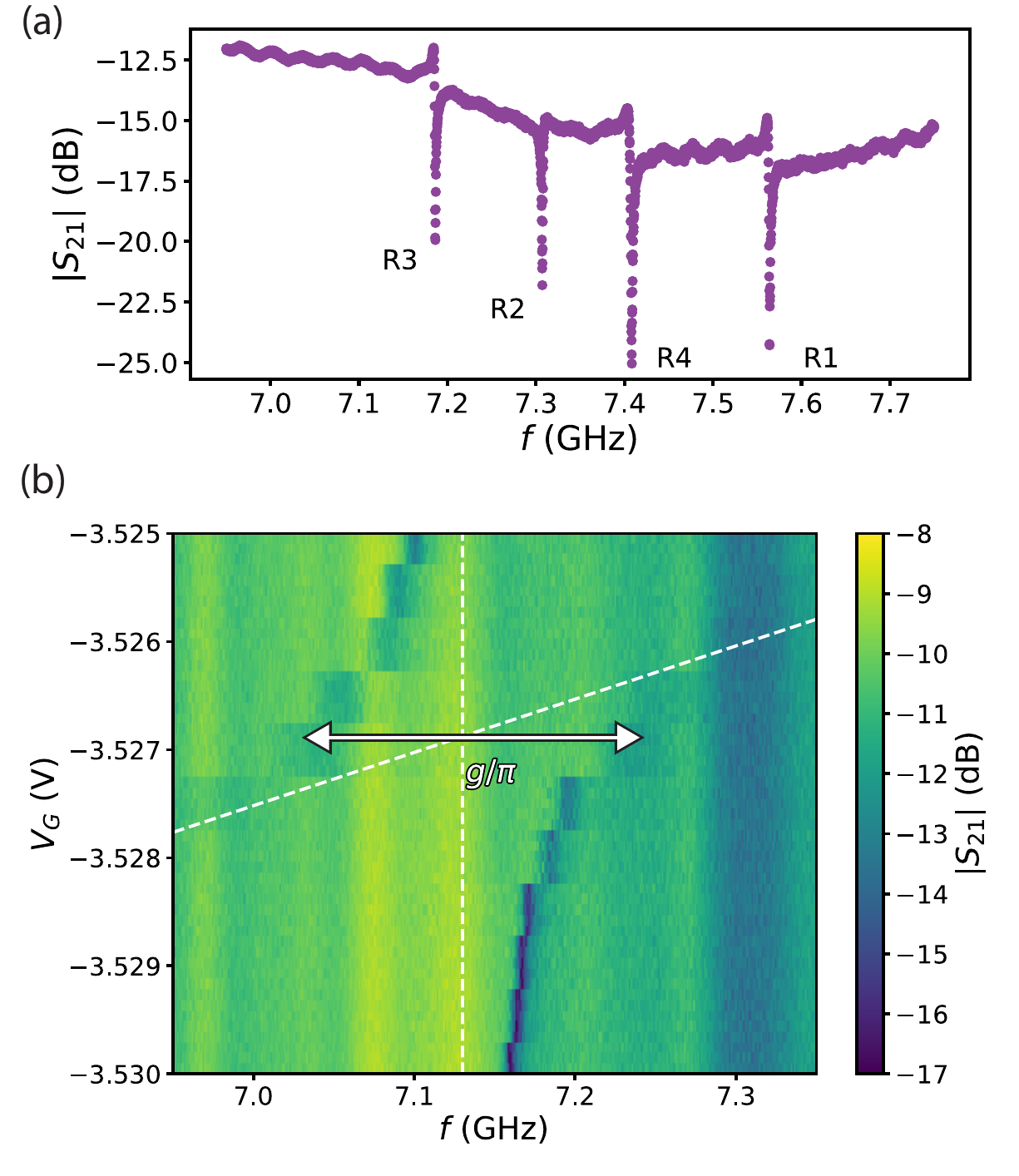}
    \caption{\textbf{Continuous wave measurement} (a) Measuring $|S_{21}|$ across the transmission line we find absorption at four distinct frequencies corresponding to the resonant frequencies of the readout resonators. (b) The junction gate voltage $V_G$ tunes the qubit frequency, and near the readout resonator frequency a vacuum Rabi splitting is observed. We find this corresponds to a coupling strength of $g/2\pi$ = 95 MHz.}
    \label{fig:2}
\end{figure}

\begin{figure}
    \centering
    \includegraphics[width=.45\textwidth]{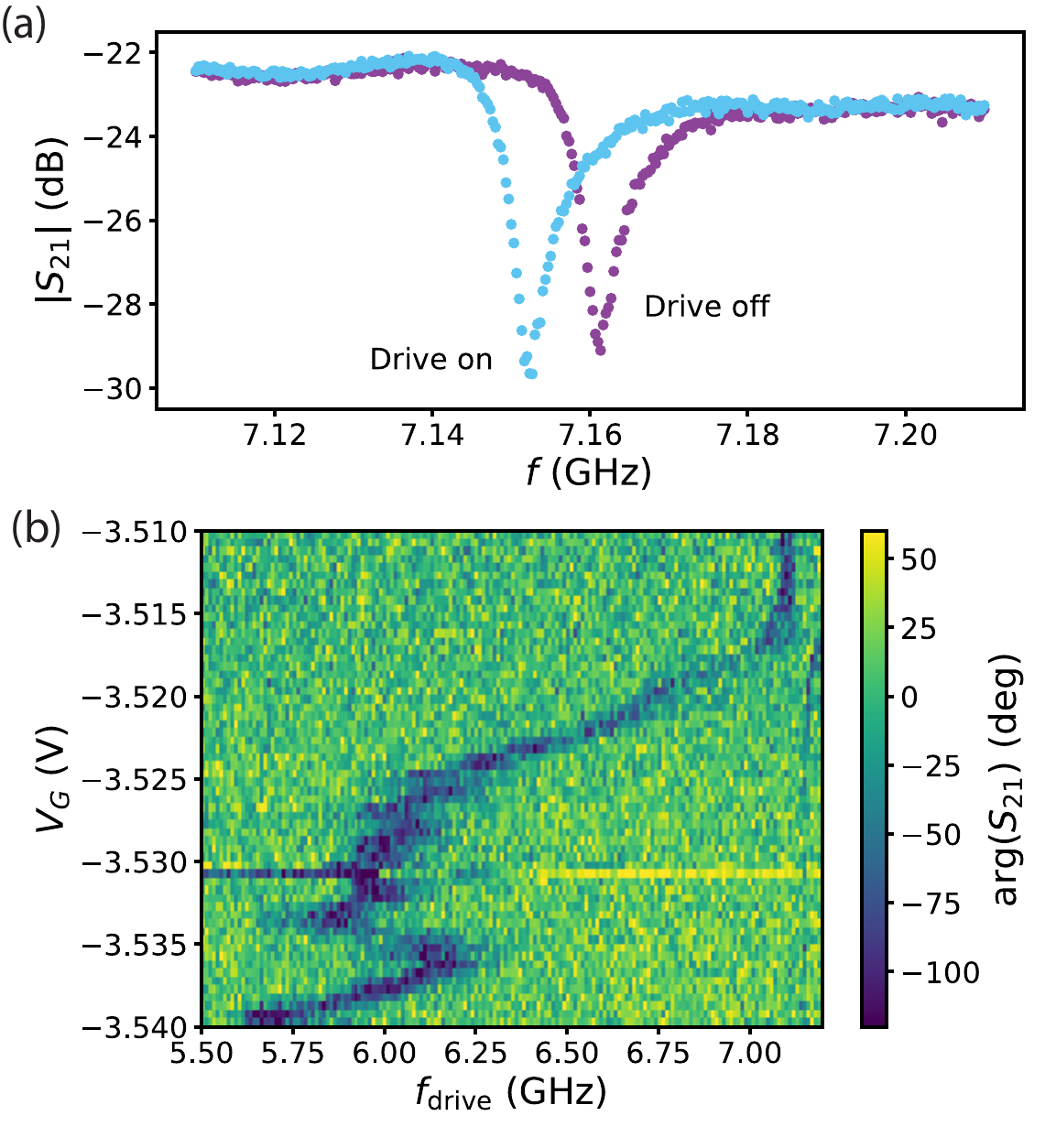}
    \caption{\textbf{Two tone spectroscopy} (a) Applying a drive tone on the qubit at a frequency of $f_Q$ = 6.55 GHz, detuned from the readout resonator by $\Delta$ = 610 MHz, the readout resonator dispersively shifts when the drive is turned on. (b) Sweeping the top gate voltage $V_G$ and applying a drive tone with varying frequency $f_\mathrm{drive}$, we find that two-tone spectroscopy reveals the qubit response to gate voltage.}
    \label{fig:3}
\end{figure}

\section{Qubit measurements}

A detailed outline of the fabrication procedure can be found in Fig. 1 and further details are outlined in Appendix A. The InAs quantum well is grown by molecular beam epitaxy and capped with a 30 nm Al layer \textit{in-situ}. Patterning is done by electron beam lithography and chemical wet etching defines the circuit. The qubit chip consists of four isolated qubits, each with a readout resonator, drive line, and gate. the resonators are coupled to a common feedline. Details of the device design can be found in Appendix B. 

Measuring the complex transmission $|S_{21}|$ across the feedline as a function of frequency, four sharp dips in magnitude at distinct frequencies can be seen, as shown in Fig. \ref{fig:2}(a), corresponding to each of the four readout resonators on the chip labelled according to their placement on the chip from left to right. We focus our study on the qubit coupled to resonator R3. Sweeping the top gate voltage $V_G$, and measuring the transmisssion near the frequency of resonator R3, we find that the qubit and readout resonator modes exhibit a vacuum Rabi splitting, where the splitting in frequency is equal to $2g/2\pi$. This can be seen in Fig. \ref{fig:2}(b) where the minimum detuning of the two modes is noted. The bare frequencies of the two modes as a function of gate voltage are shown as white dashed lines. Extracting the measured coupling strength we find that $g/2\pi = \SI{95}{\mega Hz}$, within 15\% of the value expected from finite element simulations.

We then perform dispersive measurements of the qubit state \cite{blais2004}. In a coupled qubit-resonator system in the dispersive regime, with large coupling and detuning $g/\Delta \ll 1$, we expect the readout resonator frequency to have a qubit state dependent frequency, $f_r \pm \chi$, where $\chi$ is the dispersive shift. Thus, for a sufficiently small linewidth $\kappa \lesssim \chi$, a measurement of the transmission through the resonator would allow for the unique determination of the qubit state. 

The qubit is detuned from the readout resonator by applying a gate voltage of $V_G = \SI{-3.535}{V}$ and the transmission across the feedline at the resonant frequency of R3. Sweeping the frequency of the drive tone at a power of $P_\mathrm{drive}$ = -36 dBm, we find that at a frequency $f_\mathrm{drive} = \SI{5.62}{\giga Hz}$, the readout resonator exhibits a shift down in frequency corresponding to the excitation of the qubit $|0\rangle$ to $|1\rangle$ transition. This can be seen in Fig. \ref{fig:3}(a), where we measure a shift of  $\SI{8.66}{\mega Hz}$ between the resonator frequency when the drive is on. The continuous microwave tone incoherently drives the qubit $|0\rangle$ to $|1\rangle$ transition, therefore shifting the readout resonator by less than $2 \chi$.

We then repeat this measurement while sweeping the gate voltage, modifying the current through the junction and thus the qubit frequency. As shown in Fig. \ref{fig:3}, we find that the observed qubit frequency generally decreases with gate voltage, which is expected for decreasing critical current with decreasing gate voltage. We find wide tunability $>$ \SI{1}{\giga Hz} of the qubit frequency via gate voltage. Furthermore, the qubit response to gate voltage is non-monotonic and is similar to what has been observed in the past for gatemon qubits in Refs. \citenum{Casparis2018, Larsen_PRL, danilenko2022, deLange2015, hertel2022}. It was discussed in Ref. \citenum{danilenko2022} that in an InAs nanowire device, at very small junction critical currents on the order of 10 nA, the junction becomes subject to universal conductance fluctuations.

A two-level quantum system undergoes Rabi oscillations when driven at the transition frequency between the two levels, where the final qubit state depends on the width of the drive pulse $\tau_\mathrm{Rabi}$. We measure the dynamics of the qubit in the time domain by coherently driving Rabi oscillations and measuring the characteristic lifetime. 

\begin{figure}
    \centering
    \includegraphics[width=.45\textwidth]{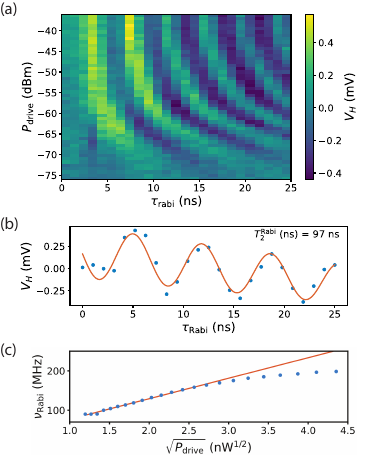}
    \caption{\textbf{Rabi oscillations:} (a) Homodyne detection voltage $V_H$ as a function of Rabi pulse width $\tau_\mathrm{Rabi}$ with the drive power $P_\mathrm{drive}$ varied. (b) Fitting a linecut of the data in (a) at a power of -52.5 dBm to a decaying sinusoid. We find a time constant $T_2^\mathrm{Rabi} = \SI{97}{\nano s}$ characterizing the coherence. (c) Extracting Rabi oscillation frequency $\nu_\mathrm{Rabi}$ at different drive powers, we find at low power that the data fits roughly to $\sqrt{P_\mathrm{drive}}$. At high power, the data deviates from the expected square root dependence, possibly due to high power nonlinearities.}
    \label{fig:4}
\end{figure}

Shown in Fig. \ref{fig:4}(a) is the homodyne detection voltage measured as pulses of varying widths are sent to the drive line of the qubit. The qubit is set to a frequency of $f_{01} = $ \SI{6.56}{\giga Hz} with the gate voltage set to $V_G$ = -3.530 V. As we vary the drive pulse width, $V_H$ oscillates, periodic in the pulse width $\tau_\mathrm{Rabi}$. We find that the frequency of these oscillations decreases with decreasing power, as is expected for Rabi oscillations. Fitting the data at $P_\mathrm{drive}$ = -52.5 dBm with the method of least squares, we extract a time constant of $T_2^\mathrm{Rabi} = \SI{97}{\nano s}$. There are four free parameters in the fits to Rabi oscillations: the time constant characterizing the exponential decay $T_2^\mathrm{Rabi}$, the oscillation frequency, and the slope and y-intercept of the decaying contribution. This linear contribution to the signal $V_H$ was identified in Ref. \citenum{hertel2022} and could possibly be due to leakage into higher levels \cite{peterer2015}. In Fig. \ref{fig:4}(c) we plot the extracted Rabi frequency in blue markers versus the square root of the drive power. It is seen that at the low power regime, the frequency follows $\sqrt{P_\mathrm{drive}}$, shown as an orange line.

\begin{figure}
    \centering
    \includegraphics[width=0.45\textwidth]{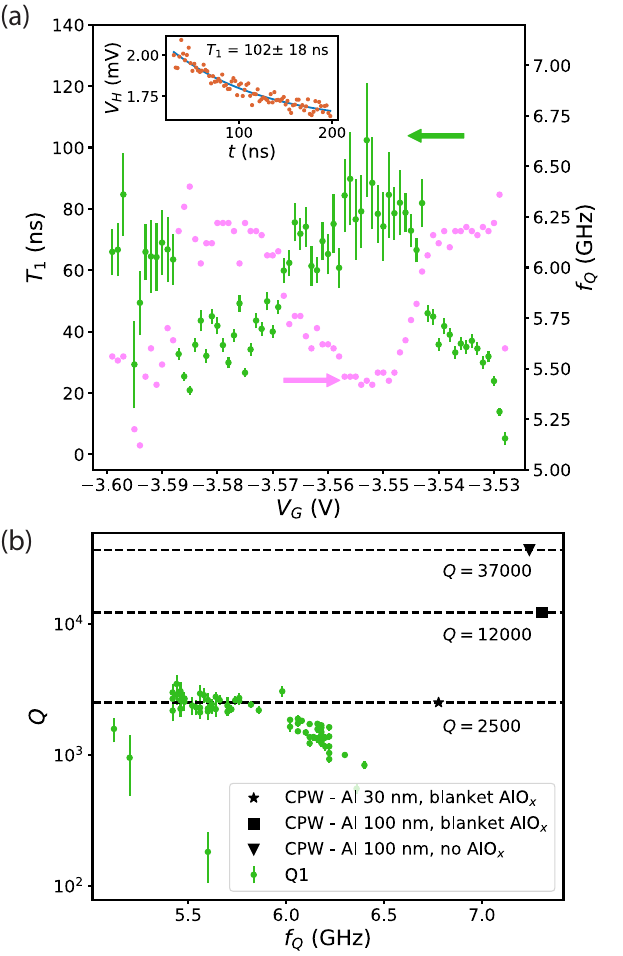}
    \caption{\textbf{$T_1$ measurements} (a)  Applying a pi pulse and measuring in time $t$ we find that the decay of the qubit state can be fit to an exponential to extract the time constant for qubit decay $T_1$ = 102 ns. The $T_1$ measurement is repeated over a range of gate voltages. (b) Qubit $T_1$'s (green) and frequencies (pink) as a function of gate voltage. }
    \label{fig:5}
\end{figure}

By calibrating the pulse width to half a Rabi period, the qubit is coherently driven to the $|1\rangle$ state. By then measuring the decay to the $|0\rangle$ state averaged over many runs, the characteristic time of this decay $T_1$ can be extracted. At a qubit frequency of $f_Q = \SI{6.51}{\giga Hz}$, we apply a \SI{10}{\nano s} wide pulse with a drive power of $P_\mathrm{drive} = \SI{-41}{dBm}$ and average over \SI{2e5}{} runs. The decay of the measured signal $V_H$ is fit to a decaying exponential, and we extract a time constant of $T_1 = \SI{102}{} \pm \SI{18}{\nano s}$ as shown in the inset of Fig~\ref{fig:5}(a). 

We measure $T_1$ over a range of gate voltages and qubit frequencies. As shown in Fig. \ref{fig:5}(a), the qubit frequency $f_Q$ again exhibits a non-monotonic tuning between \SI{6.5}{\giga Hz} and \SI{5.2}{\giga Hz} over the gate voltage range $\SI{-3.545}{V} < V_G < \SI{-3.535}{V}$, similar to the two-tone measurement in Fig. \ref{fig:3}(b). We find that the measured $T_1$ also vary over this range, generally increasing with decreasing qubit frequency. In order to understand further the spread of $T_1$ versus frequency we measure CPW resonators as a proxy for qubit lifetime. 

\section{Loss Mechanisms}

Qubit and resonator loss can be decomposed into a sum of contributions from various loss mechanisms, including dielectric, inductive, quasiparticle, and radiative losses \cite{mcrae2020}. In the following section we detail a systematic study of loss in CPW resonators in order to understand the role of inductive and dielectric loss in our qubit devices. 

We consider a loss model which takes into account inductive and dielectric losses
\begin{equation}
    \frac{1}{Q_\mathrm{\rm int}} = \sum_i p_i \tan\delta_i + \alpha\frac{1}{Q_S}+ \frac{1}{Q_0}.
\end{equation}
The first term is due to dielectric loss, where for a given interface or volume $i$, $p_i$ is the participation ratio of that layer, being the integral of the electric field squared in each layer normalized to the integral over all space of the electric field squared. The dielectric loss of that layer is $\tan\delta_i$. The second term characterizes inductive losses. The prefactor $\alpha$ is the kinetic inductance fraction which is related to the London penetration depth $\lambda$ and the ratio of the surface to volume magnetic field energy $p_\mathrm{mag}$ through $\alpha = \lambda p_\mathrm{mag}$. The surface quality factor $Q_S$ is related to dissipative conduction in the superconductor
\begin{equation}
    Q_S = \frac{\omega \mu \lambda}{R_S} = \frac{X_S}{R_S},
\end{equation} where $R_S$, the surface resistance, and $X_S$, the surface reactance,are the real and imaginary parts of the surface impedance respectively. These quantities are both functions of the temperature $T$. This loss model is applied to measurements of the internal quality factor $Q_\mathrm{int}$ of CPW resonators. Six resonators are fabricated on each chip in a hanger styel with a common feedline. In certain devices some resonators are unable to be identified in large range transmission measurements due to errors in fabrication, such as the presence of scratches or contaminants. Fig. 6(a) and (b) show the magnitude and phase of $S_{21}$ at a frequency near the resonant frequency of R6 from the 130 nm sample. Fits are conducted using the circle fitting method detailed in Ref. \citenum{Probst2015}.

\begin{figure}
    \centering
    \includegraphics[width=.5\textwidth]{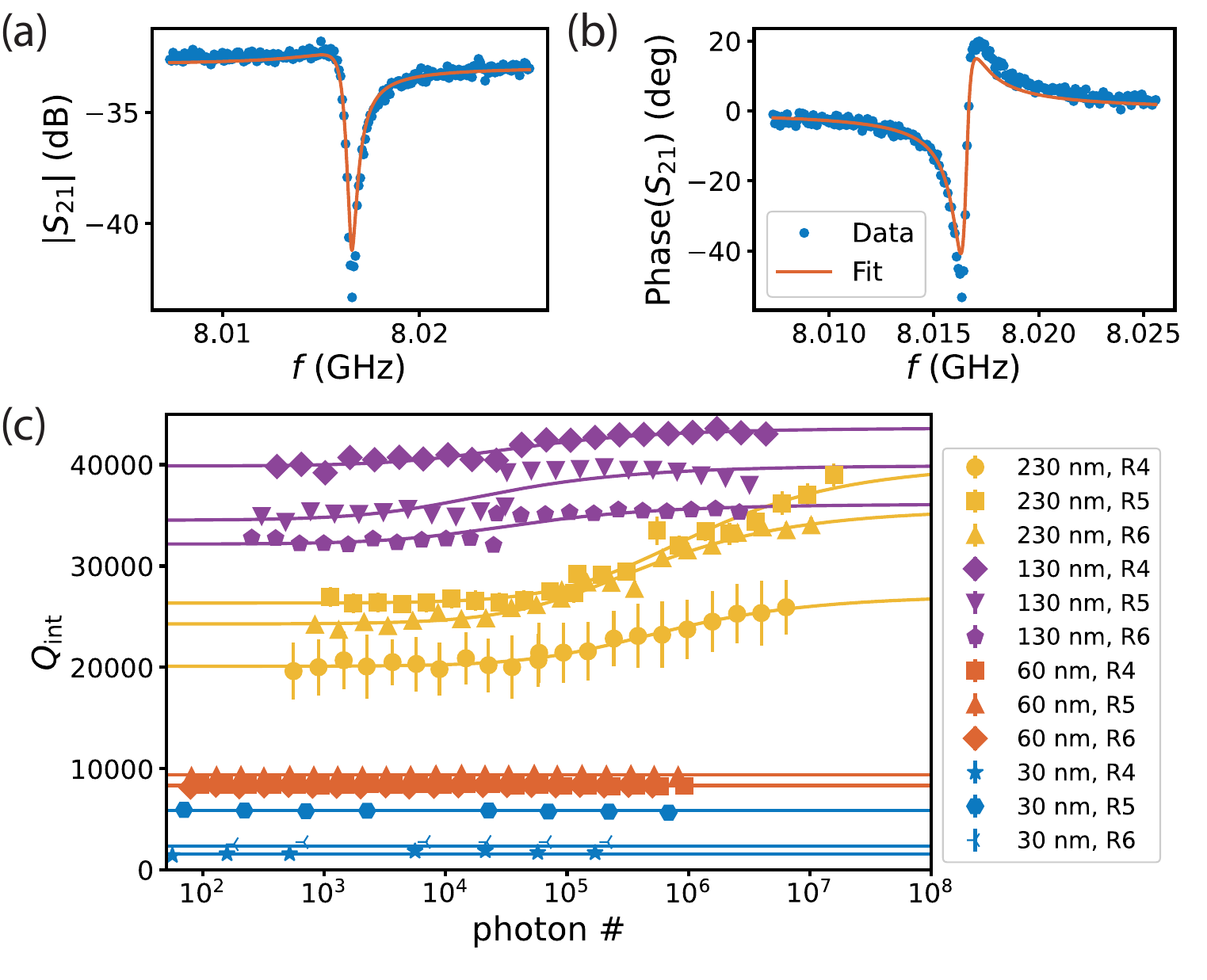}
    \caption{\textbf{Internal quality factors and power dependence}. Amplitude and phase of the transmission $S_{21}$ are shown in (a) and (b) as well as fits shown as solid lines. From the fits we are able to extract $f_r$ and $Q_\mathrm{int}$. (c)Internal quality factors vs. number of photons in the cavity for resonator devices with varying Al thicknesses. Devices on thick Al films (130 and 230 nm) can be seen to exhibit some power dependence, and are fit to a model of loss following two-level systems (solid lines), while thin Al films (30 and 60 nm) show power independent loss, suggesting that the total loss is dominated by power-independent loss mechanisms, such as inductive loss.}
    \label{fig:qualityfactors}
\end{figure}

\begin{figure}
    \centering
    \includegraphics[width=0.5\textwidth]{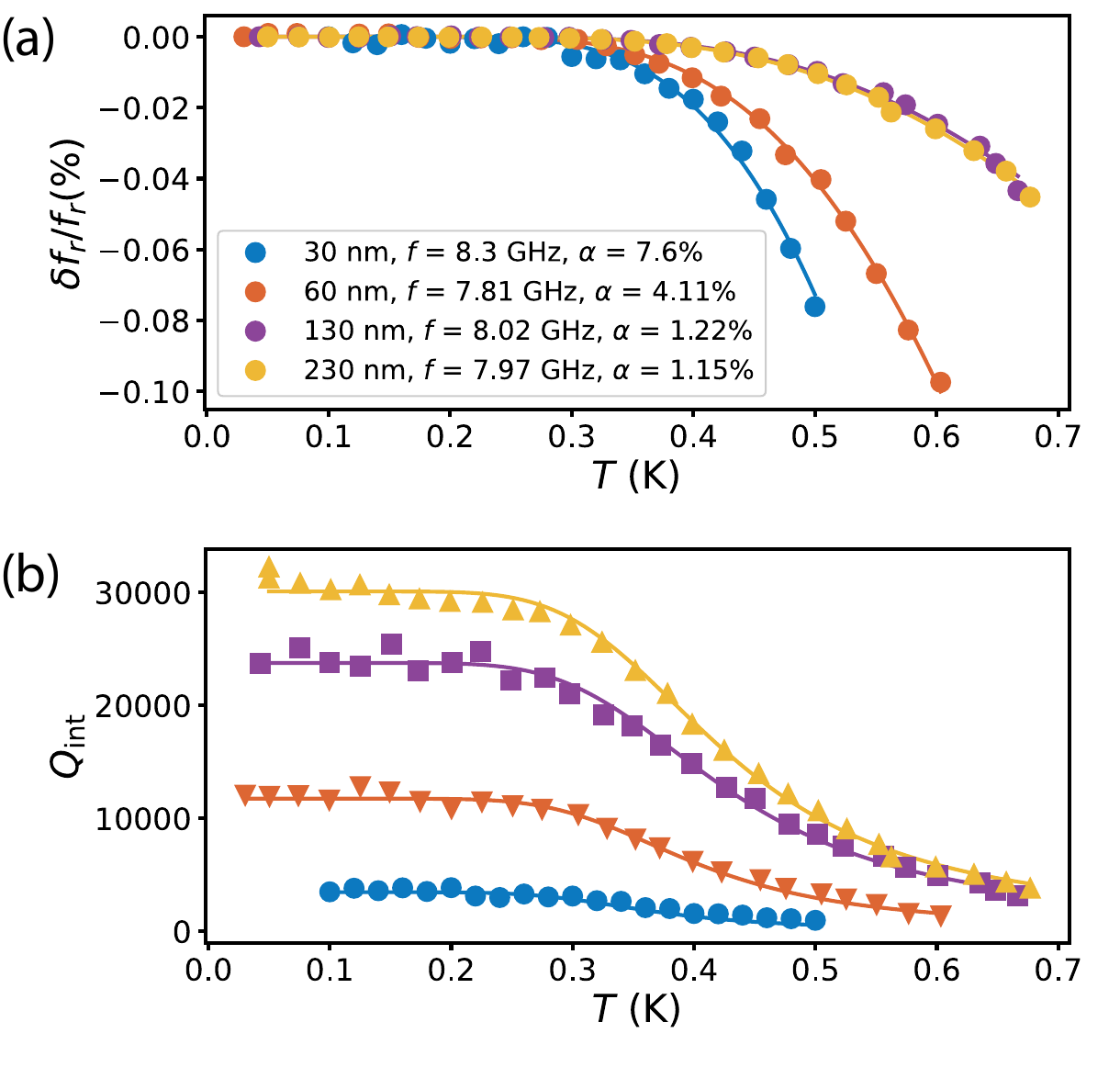}
    \caption{\textbf{Temperature dependence.} Percent change in frequency in (a) and  $Q_\mathrm{internal}$ in (b) as a function of temperature with fits to the Mattis-Bardeen model (solid lines) for the four different Al thicknesses.}
    \label{fig:QvsT}
\end{figure}



\subsection{Dielectric Losses}

Two-level systems (TLSs) have been known to limit device performance in many high coherence qubits, and the microscopic origin of TLSs in superconducting devices is an active area of research. We fit power dependence of the quality factor to the non-interacting TLS model \cite{mcrae2021}

\begin{equation}
    \frac{1}{Q_\mathrm{int}(n)} = \frac{1}{Q_0} + \frac{1}{Q_\mathrm{TLS}} \frac{\tanh\left(\frac{\hbar \omega}{2 k_B T}\right)}{\sqrt{1+\frac{n}{n_c}}}.
\end{equation}
Here $Q_0$ takes into account power independent loss mechanisms, $1/Q_\mathrm{TLS}$ is the associated TLS loss, and the ratio $n/n_c$ is the average number of photons in the cavity over some critical photon number at which TLSs start to get saturated.

The results are shown in Fig. 5, where we include data from three devices (R4, R5, and R6) from four different samples (Al thicknesses of 30, 60, 130, and 230 nm). We find that, the samples with thicker Al have larger internal quality factors, consistent with the expectation that thicker films mitigate inductive loss, as will be expanded on in the next subsection. We also find that in samples with 130 and 230 nm thick Al, $Q_\mathrm{int}$ decreases with the number of photons in the cavity, which is consistent with the expectation from two level system loss. By fitting this data to Equation 3, we are able to extract $Q_0$, $Q_\mathrm{TLS}$, and $n_c$. These results are featured in Table 1. Samples with thinner Al show power independent loss, which indicates that other power-independent loss mechanisms, such as inductive loss, dominate the total measured loss. The plotted traces for the 30 and 60 nm thick films are simply the average $Q_\mathrm{int}$ measured over the whole power range.

We find that the samples with 130 and 230 nm Al have $Q_0$ in the range of 2.7 to \SI{4.4e4}{}. This is consistent with loss measurements on GaAs substrates, where the limits to coherence were shown to be due to the piezoelectricity of III-V substrates \cite{Scigliuzzo_2020, mcrae2020}. The values extracted for $Q_\mathrm{TLS}$ varied between the samples, being around 1.4 to \SI{2.4e3}{} for the 130 nm Al sample and 4.0 to \SI{6.0e2}{} for the 230 nm sample. The lower $Q_\mathrm{TLS}$ observed for the 230 nm sample is possibly due to insufficient cleaning of the sample surface after the III-V buffer layer etch.

\subsection{Inductive Losses}

As the temperature is varied, the real and imaginary components of surface impedance cause changes to the resonant frequency and quality factor of a CPW resonator \cite{turneaure_surface_1991}. 
The complex impedance $Z_S = R_S+j \delta X_S$ follows
\begin{equation}
    Z_S(T) = \frac{\omega\mu_0\lambda}{p_\mathrm{mag}} \left(\frac{1}{Q} + 2j\frac{\delta f}{f}\right).
\end{equation} 
We apply a similar procedure to fit temperature dependence data as is described in Refs. \citenum{Gao2008, elfeky2023quasiparticle}, and a description of the fitting procedure can be found in Ref. \citenum{elfeky2023quasiparticle}.

\begin{table*}[tbp!]
\centering
\begin{tabular}{|| c | c | c | c | c | c | c | c | c | c ||} 
 \hline
 Al thickness & Resonator & $f_r$ (GHz) & $Q_0$ ($\times 10^3$) & $Q_\mathrm{TLS}$ ($\times 10^3$) & $n_c$ ($\times$ \SI{e4}{}) & $\alpha$ (\%) & $L_k^\square$ (pH)& $\lambda$ (nm) \\ [0.5ex] 
 \hline\hline

 30 nm & - & 8.304 & 3.48 & - & - & 7.60 & 1.2248 & 164 \\ 
 \hline
  \multirow{5}{*}{30 nm}
  & R1 & 5.835 & \SI{7.45}{} & - & - & - & -& - \\ 
  & R2 & 6.107 & \SI{1.70}{} & - & - & - & -& - \\ 
  & R3 & 6.306 & \SI{1.55}{} & - & - & - & -& - \\ 
  & R4 & 6.737 & \SI{5.87}{} & - & - & - & -& - \\ 
  & R6 & 7.412 &  \SI{2.36}{} & - & - & - & -& - \\ 
 \hline
   \multirow{5}{*}{60 nm}
  & R1 & 6.140  & \SI{7.52}{} & - & - & 5.29 & 0.853 & 137 \\ 
  & R2 & 6.396  & \SI{8.37}{} & - & - & 4.24 & 0.683 & 122 \\ 
  & R4 & 7.061  & \SI{8.39}{} & - & - & 4.85 & 0.782 & 131 \\ 
  & R5 & 7.413  & \SI{9.39}{} & - & - & 4.24 & 0.683 & 122 \\ 
  & R6 & 7.808  & \SI{8.30}{} & - & - & 4.69 & 0.756 & 129 \\ 
 \hline
 \multirow{3}{*}{130 nm}
  & R4 & 7.244 & \SI{43.6}{} & \SI{2.43}{}& \SI{1.30}{} & 1.29 &\SI{0.208}{} & 66\\ 
  & R5 & 7.605 & \SI{39.9}{} & \SI{1.40}{}& \SI{0.733}{} & 1.30 & \SI{0.210}{} & 66 \\ 
  & R6 & 8.017 & \SI{36.1}{} & \SI{1.70}{}& \SI{1.20}{} & 1.36 & \SI{0.219}{} & 68 \\ 
 \hline
 \multirow{5}{*}{230 nm}
  & R1 & 6.278 & \SI{31.5}{} & \SI{0.547}{} & \SI{13.0}{} & 1.24 & \SI{0.200}{} & 65 \\ 
  & R2 & 6.551 & \SI{34.8}{} & \SI{0.606}{} & \SI{12.9}{} & 1.39 & \SI{0.224}{} & 69 \\ 
  & R4 & 7.215 & \SI{27.1}{} & \SI{0.404}{} & \SI{14.7}{} & 1.38 & \SI{0.222}{} & 69\\ 
  & R5 & 7.587 & \SI{40.0}{} & \SI{0.421}{} & \SI{22.5}{} & 1.32 & \SI{0.213}{} & 67\\ 
  & R6 & 7.972 & \SI{35.6}{} & \SI{0.438}{} & \SI{8.83}{} & 1.31 & \SI{0.211}{} & 67 \\ 
 \hline
\end{tabular}
\caption{Summary of the results for CPW resonators with varying thicknesses of Al. On each chip the resonators identified in a broad frequency scan are included. The Al thickness denotes the sample, being 30 nm of epitaxial Al plus an additional thickness of sputtered Al. The measured frequency $f_r$ are shown for the given devices indexed as $R1$ for the lowest frequency resonator by design, for example. Fits to the TLS model allow us to extract the power independent quality factor $Q_0$, the quality factor associated with two level system loss $Q_\mathrm{TLS}$, as well as the critical photon number $n_c$. Fits to the Mattis-Bardeen theory for the temperature dependence of $f_r$ allow us to extract the kinetic inductance fraction $\alpha$, and subsequently the London penetration depth $\lambda$.}
\label{table:2}
\end{table*}

Using this equation for the fractional frequency shift it is possible to fit the data as a function of temperature in order to extract the kinetic inductance fraction $\alpha$ and the superconducting gap $\Delta$.

We measure $S_{21}$ at varying temperatures and study the resonant frequency and the quality factor change as a function of temperature. The data was taken with the sample at the base temperature of \SI{15}{\milli K} and increased to a maximum temperature of \SI{800}{\milli K}. Before the data was taken at each temperature, a 30 minute dwell time was applied to allow the temperature to equilibriate. The results can be seen in Fig. 7 (a) and (b). We show one resonator (R6) from three devices with varying Al film thicknesses: 60, 130 and 230 nm. Data from a fourth CPW is included with a different resonant frequency and coupling to the feedline, but identical center width and gap. We then fit the change in resonant frequency to the Mattis-Bardeen theory detailing the change in the complex surface impedance as a function of temperature. It can be seen that the measured $Q_\mathrm{int}$ of R6 for the 130 nm thick Al sample is lower than that in Fig. 6(c), possibly because the temperature dependence measurement was conducted on a different cooldown, and between the power and temperature sweeps the sample was kept in ambient conditions for approximately one month.

In all devices a decrease in frequency with temperature is seen, consistent with the Mattis-Bardeen theory. We also find that the temperature at which the frequency starts to change decreases with decreasing film thickness. We fit each data set to the Mattis-Bardeen theory in order to extract the kinetic inductance fraction $\alpha$ while simultaneously fitting for the superconducting gap $\Delta$ where we consistently find $\Delta \approx \SI{210}{\micro eV}$ for all films, similar to what has been seen in the past for thin Al films \cite{mayer2019}. The fit results show that $\alpha$ systematically decreases as the thickness increases. For the epitaxial Al film, $\alpha = 7.6\%$. For the film 60 nm thick Al, the kinetic inductance decreases to around 4.2 to 5.3 \%. For the 130 nm film and the 230 nm film the kinetic inductance saturates at around 1.2 to 1.4 \%. Multiple resonators on the same sample show similar results for $\alpha$ to within 20\% of each other for the 60 nm sample and within 10\% of each other for the 130 and 230 nm samples.

Most notable are the measured internal quality factors versus Al film thickness. CPWs on the thinnest Al film had the lowest measured $Q_\mathrm{int} = \SI{3.5e3}{}$. The sample with 60 nm thick Al had internal quality factors between 1.6 and \SI{7.5e3}{}, while the 130 nm sample had $Q_\mathrm{int}$ in the range of 35 to \SI{45e3}. The 230 nm sample showed similar $Q_\mathrm{int}$ to that of the 130 nm sample between \SI{27}{} and \SI{40e3}{}. The results are summarized in Table 1. 

As the film thickness of aluminum decreases, the London penetration depth increases from it's bulk value of 50 nm \cite{deGennes}. In the regime where the film thickness becomes less than the London penetration depth, the kinetic inductance becomes significantly enhanced, as seen in our devices. As detailed in Refs. \citenum{kautz2008, lópeznúñez2023magnetic}, one is able to relate the kinetic inductance per square $L_k^\square$ of a thin film to the London penetration depth through 

\begin{equation}
    L_k^\square = \mu_0 \lambda \coth\left(\frac{d}{\lambda}\right),
\end{equation}

where $d$ is the film thickness. With an accurate measure $\alpha$ and the characteristic impedance of the CPW we are able to find $L_k^\square$ and thus the London penetration depth. The results are shown in Table 1. We find that for the 230 and 130 nm films, the London penetration depths are close to the expected bulk value of aluminum, being between 65 and 70 nm for all devices. For the 60 nm sample, the penetration depth increases to around 125 nm, making $\lambda>d$ for this film. Finally, for the 30 nm Al film, we find that the penetration depth is 160 nm, even further in $\lambda>d$ regime, with $\lambda$ more than five times larger than the film thickness.

\section{Discussion}
%

In this section we compare the measured loss of CPWs to the gatemon qubit presented in the earlier section of this article. In Fig. 5(b) we plot the measured $T_1$ values in terms of an equivalent qubit quality factor, $Q = 2\pi f_Q T_1$, as a function of qubit frequency $f_Q$. We compare the measured qubit quality factors to low power measurements of $Q_\mathrm{int}$ of the CPWs. It is clearly seen that the observed qubit $Q$ is lower than all measured CPWs, and closest to $Q_\mathrm{int}$ measured for the 30 nm thick Al sample. 

We believe there are two main differences in the fabrication procedure of the highest $Q$ CPW resonators and that of gatemon qubits that decrease the measured qubit $Q$. The first reason is the thin 30 nm Al used in the gatemon qubit, which as we have shown, could introduce a considerable amount of inductive loss. The second difference is the use of a blanket AlOx layer, which is known to have loss tangents in the $\tan{\delta}\sim 1\times10^3$ range. In a separate measurement we found that a sample with thick Al and blanket AlOx had $Q$ reduced to around $1.2\times 10^4$. This suggests that the limit set by inductive loss of a thin Al film dominates the observed qubit loss, but is enough substantially decrease the $Q$ of CPWs with thick Al. 

\begin{figure}
    \centering
    \includegraphics[width=0.45\textwidth]{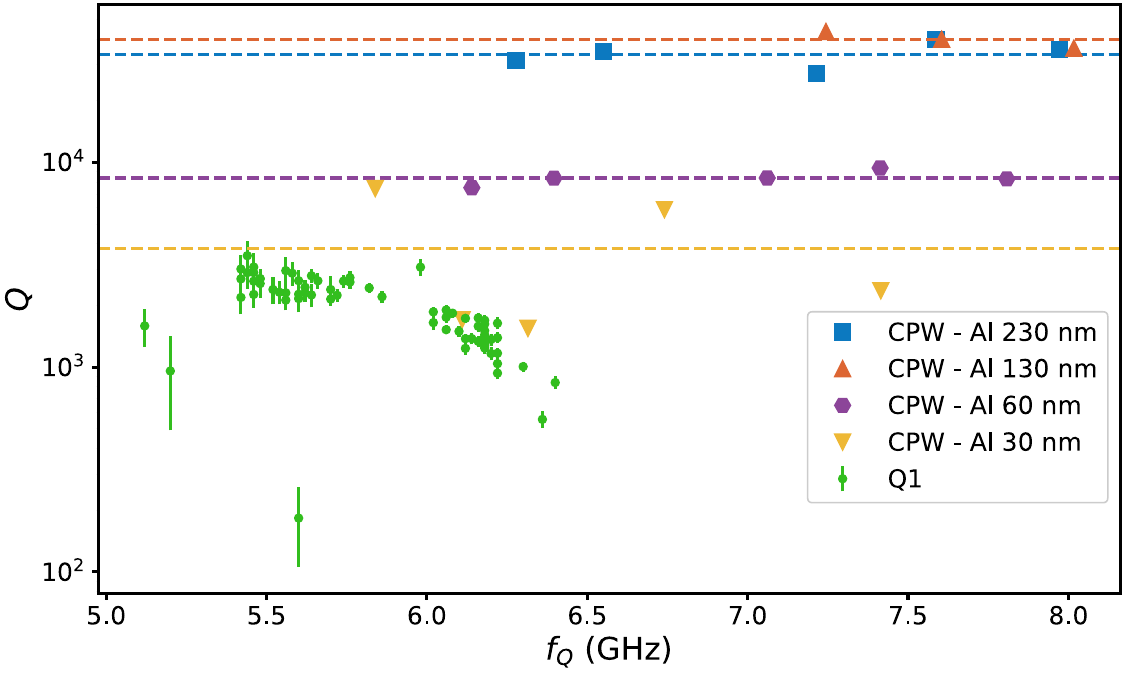}
    \caption{\textbf{Quality factors of CPW resonators and qubits:} The qubit quality factor $Q = 2\pi f_Q T_1$ and low power resonator quality factor is plotted versus frequency. Dashed lines are average quality factor over all of each samples on each chip.}
    \label{fig:QvsT}
\end{figure}

The CPW measurements presented here detail immediate next steps to enhance gatemon relaxation times by reducing inductive and capacitive losses. Dielectric loss from the blanket AlO$_x$ layer can be reduced by patterning and lifting off the gate dielectric, or by using hexagonal boron nitride, a low-loss, small form factor gate dielectric \cite{Barati2021, Wang2022}. In order to reduce inductive losses, a thicker \textit{in-situ} Al layer can be deposited, or a thick, low-loss superconducting layer can be deposited \textit{ex-situ} via sputtering, given sufficient cleaning before the deposition. Decreasing inductive losses in the superconducting film will also manifest as decreased Purcell loss through the readout resonator. We note that inductive loss in the junction, radiative loss through the gate line, and quasiparticle loss may also play a role in limiting qubit quality factors \cite{elfeky2023quasiparticle}.


We believe that further surface preparation optimization can be done in order to reduce TLS loss in future devices. It has been shown in the past that various etching recipes yield different device performance results \cite{Earnest_2018}. We plan to investigate different types of acid etching in order to passivate the surfaces both before deposition and during the patterning of the devices. It is also unclear what the effect of the ion milling is on the sample surface, and whether a shorter or longer ion mill will lead to higher quality factor devices. 

In making devices with coherence times greater than \SI{1}{\micro s}, it may be necessary to develop new techniques to reduce the participation in the III-V layers. These can include flip-chip techniques\cite{hazard2022}, epitaxial liftoff \cite{Cheng2013} and wafer bonding to a low loss substrate such as silicon or sapphire, as well as growing III-V semiconductors directly on silicon \cite{kroemer2004}.

\section{Acknowledgements}
We thank Joe Yuan, Matthieu Dartiailh, Nick Materise, and Noah Goss for fruitful conversations. We acknowledge support from the Army Research Office agreements W911NF2110303 and W911NF2210048. We also acknowledge support from MURI ONR award no. N00014-22-1-2764 P00001. W. M. S. acknowledges funding from the ARO/LPS QuaCR Graduate Fellowship.
\newpage

\begin{figure*}
    \centering
    \includegraphics[width=\textwidth]{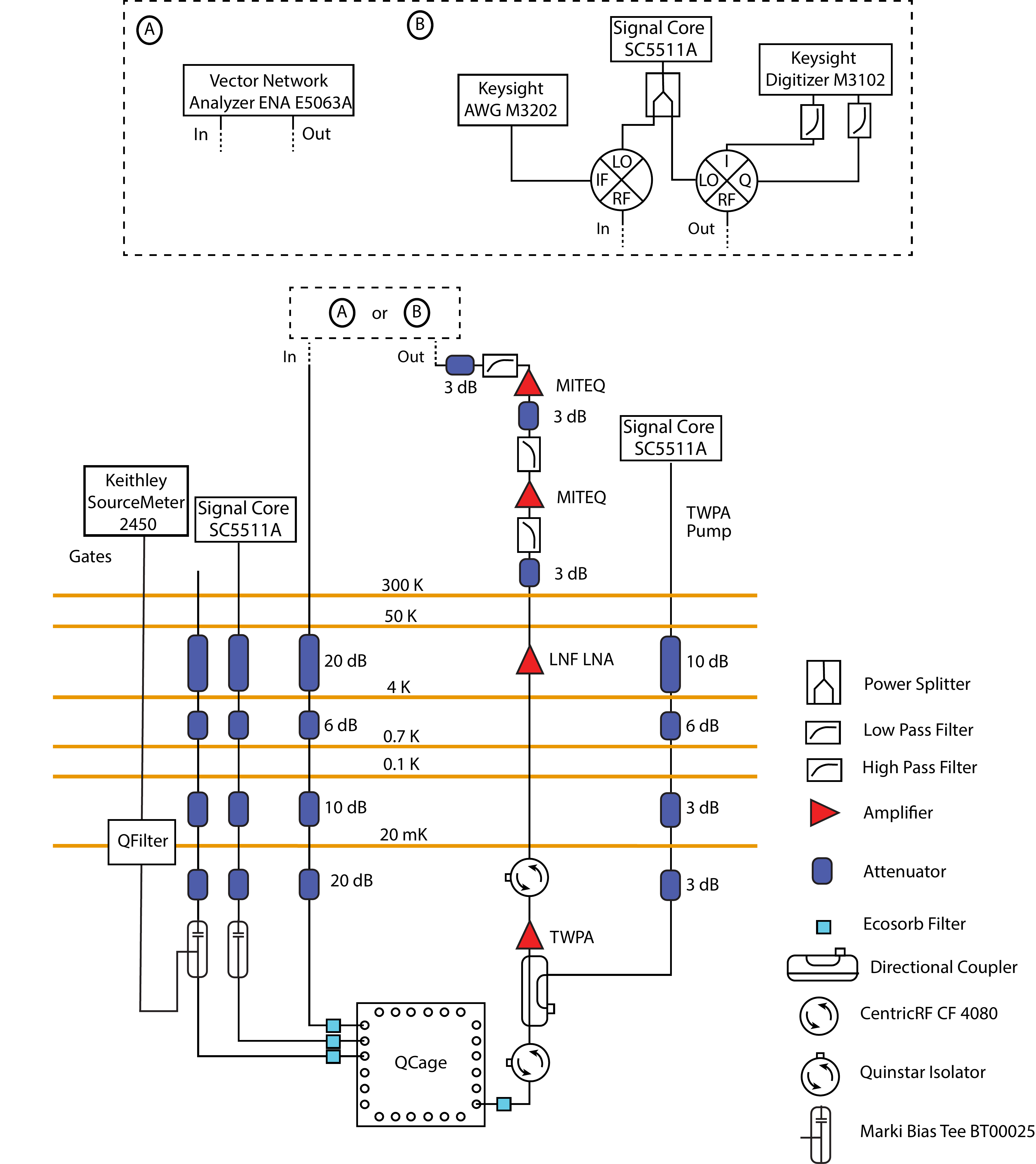}
    \caption{Measurement setup}
    \label{fig:fridge}
\end{figure*}

\section*{Appendix A: Growth and Fabrication}

The 2DEG is realized by an InAs quantum well grown near the surface by molecular beam epitaxy. A schematic of the heterostructure is shown in Fig. \ref{fig:fab}(a). An epi-ready, \SI{500}{\micro m} thick, Fe-doped, semi-insulating InP substrate is loaded into an ultra-high vacuum molecular beam epitaxy chamber. The native oxide is thermally desorbed, followed by the growth of an In$_{0.53}$Al$_{0.47}$/In$_{0.52}$Al$_{0.48}$ superlattice, a 100 nm thick In$_{0.52}$Al$_{0.48}$As layer, and a 400 nm thick In$_{x}$Al$_{1-x}$As graded buffer layer. The composition is graded from $x=0.52$ to $0.81$. The quantum well is then grown, consisting of layers of InGaAs, InAs, and InGaAs with thicknesses of 4 nm, 4 nm, and 10 nm respectively. The temperature is then lowered and a 30 nm thick layer of Al is deposited \textit{in-situ}. Details of the growth are expanded on in detail in \cite{Kaushini2018, strickland2022, Yuan2020}.

Schematics of the device fabrication process are also shown in Fig. \ref{fig:fab}, with a view of the surface shown in the left-hand panel, and a tilted, side view of the layers shown on the right panel. The process starts by dicing a $7\times7$ mm piece from the wafer. We use electron beam lithography to define the patterns and polymethyl methacrylate is used as electron beam resist. To etch the native Al layer, we use a wet chemical etchant Transene Type D, and to etch the epitaxial III-V layers, we use a solution consisting of phosphoric acid (H$_3$PO$_4$, 85\%), hydrogen peroxide (H$_2$O$_2$, 30\%), and deionized water in a volumetric ratio of 1:1:40. The first lithography and etching step defines the microwave circuit, where we etch the native aluminum layer, and the III-V layers in successive steps. The resulting pattern (to scale) can be seen in in Fig. \ref{fig:fab}(b). We next expose and etch a thin 100 nm long (separation of the two aluminum leads), \SI{5}{\micro m} wide strip to define the planar Josephson junction. This can be seen in Fig. ~\ref{fig:fab}(c), where the inset of the left panel shows a zoomed in image of the junction area (exposed semiconductor region enlarged for visibility). Following the deposition of a 40 nm blanket layer of AlO$_x$ to serve as a gate dielectric, we pattern the gates and deposit a 50 nm thick Al layer for the gate electrodes either by thermal evaporation or by sputtering. The AlO$_x$ gate dielectric can be seen in Fig.~\ref{fig:fab}(d) as an opaque white layer over the whole chip.

CPW devices are fabricated using the first lithography and etching step as described above, but do not undergo the subsequent processing steps. The CPW devices differ from each other in the fact that an additional Al layer of varying thickness was deposited on three of the samples. Sputtering is done on the MBE grown Al after an argon plasma ion milling of the sample surface at 25 W with 3 mTorr flow for 5 min prior to the deposition. We deposit Al by dc magnetron sputtering at 80 W at a rate of about 5 nm/min.

\section*{Appendix B: Qubit design}

We report on two chips containing four qubits and 6 CPW resonators respectively. On the qubit chip, each qubit is coupled capacitively to a readout resonator, drive line, and gate electrode. The readout resonators are coupled inductively to a common feedline. An optical image of one such qubit is shown in Fig.~\ref{fig:fab}(e) with an equivalent circuit diagram shown in Fig.~\ref{fig:fab}(f). We use Ansys Q3D extractor to calculate the Maxwell capacitance matrix. We find that the qubit shunt capacitance is $C_S = \SI{62.7}{\femto F}$, giving an estimated charging energy of $E_C/h = e^2/2C_S = \SI{309}{\mega Hz}$, where $e$ is the elementary charge and $h$ is Planck's constant. The Josephson junction provides a nonlinear inductance $L_J$ in parallel with a shunt capacitance. By virtue of tuning the current through the junction, the top gate electrode sets a voltage $V_G$ which controls the qubit frequency $f_Q$. At a qubit frequency of $f_Q = \SI{6}{\giga Hz}$, the qubit is detuned from the readout resonator by $> \SI{1}{\giga Hz}$, allowing for dispersive readout of the qubit state. The Josephson energy at this frequency would be $E_J = \SI{16}{\giga Hz}$ giving a ratio of $E_J/E_C = 52$, satisfying the transmon condition $E_J\gg E_C$. We note that the critical current through the Josephson junction at this frequency is $I_C = \SI{30}{\nano A}$. A $\lambda/4$ readout resonator with a frequency of $f_\mathrm{r}$ = \SI{7.14}{\giga Hz} is coupled capacitively to the qubit with an estimated coupling strength $g/2\pi = \SI{109}{\mega Hz}$. We note that the readout resonator frequency is shifted down due to an appreciable kinetic inductance of the thin film Al from a frequency of $f_\mathrm{r}^0$ = \SI{7.56}{\giga Hz} expected by design \cite{gao2006}, leading to a kinetic inductance fraction of 10\%. We will discuss the implications of this large kinetic inductance later in this article. An external drive line is coupled to the qubit by a coupling strength of $\kappa/2\pi = \SI{396}{\kilo Hz}$. 

The CPW device design consists of six $\lambda/4$ coplanar waveguide resonators inductively coupled to a common feedline. Each resonator has a varying external coupling quality factor $Q_{\mathrm{ext}}$ determined by the spacing of the resonator from the feedline $c$ and expected resonant frequency depending on the length $l$. Based on the geometry of each resonator we perform finite element simulations using Ansys \cite{ansys}. A summary of the calculated resonant frequency and external coupling quality factor are detailed in Table 1. Each coplanar waveguide has a center trace width of \SI{35}{\micro m} and a spacing to ground of \SI{20}{\micro m}.


\begin{figure}
    \centering
    \includegraphics[width=0.5\textwidth]{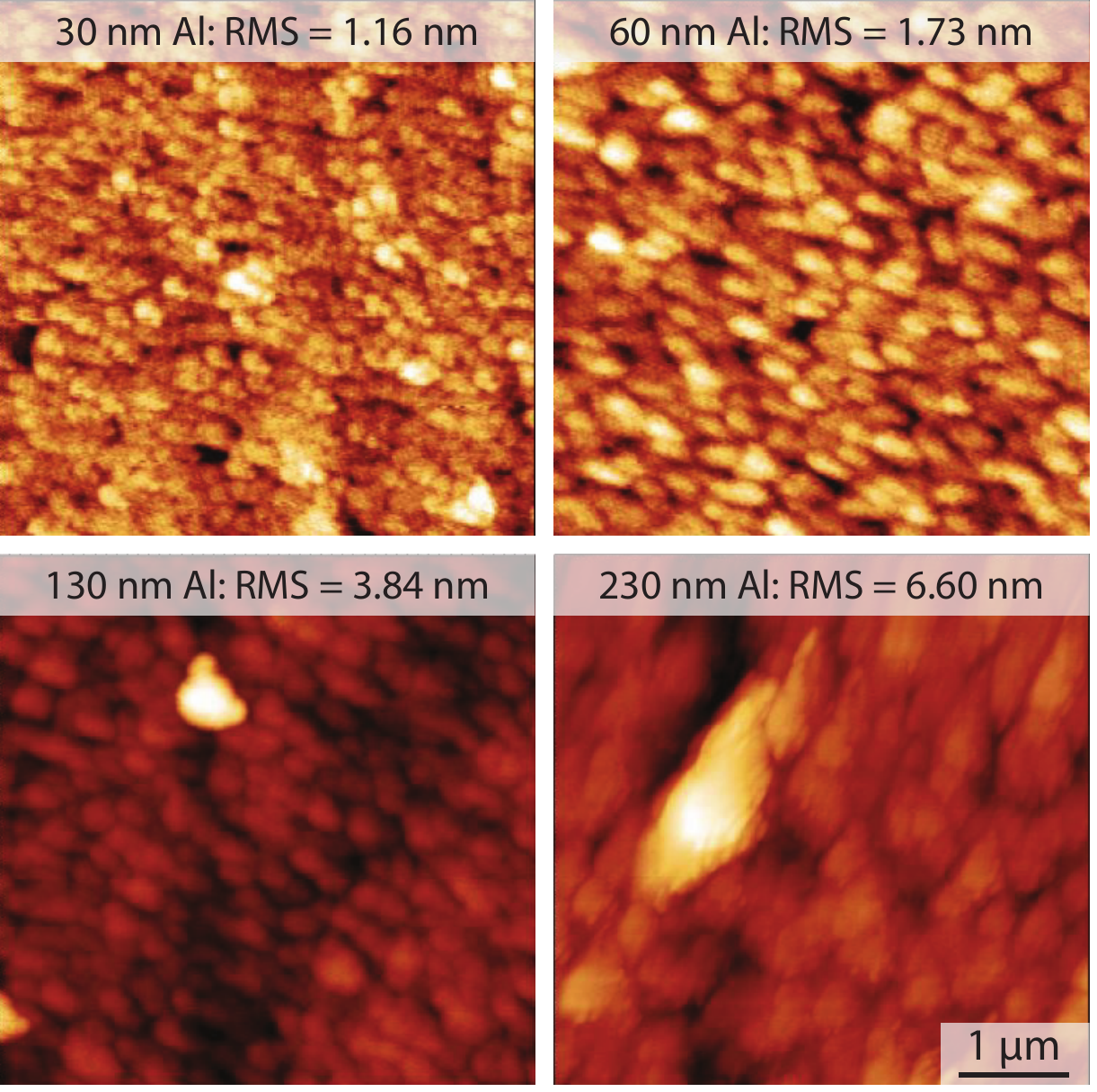}
    \caption{Atomic force micrographs of four aluminum films varying thicknesses of aluminum, being (a) 30 nm, (b) 60 nm, (c) 130 nm, and (d) 230 nm. The root-mean-squared roughness is noted for each.}
    \label{fig:enter-label}
\end{figure}

\section*{Appendix C: Measurement setup}

A schematic of the measurement setup is found in \cref{fig:fridge}. Measurements are conducted in a Triton, a croygen-free dilution refrigerator from Oxford Instruments. The device chip is placed in a QCage, a microwave sample holder from QDevil, and connected to the printed circuit board by aluminum wirebonds. RF signals are sent from a vector network analyzer or a microwave signal generator and attenuated by -56 dB with attenuation at each plate as noted. The signal then passes through an Eccosorb 
filter, made from cured castable epoxy resin. The signal is then sent through the sample, returned through another Eccosorb filter, and passed through a Quinstar isolator with 20 dB isolation and 0.2 dB insertion loss. A traveling wave parametric amplifier then amplifies the signal, which is then passed through another isolator, and then amplified with a low noise amplifier mounted to the 4K plate and two room temperature amplifiers (MITEQ) outside the fridge.

Pulsed signals are generated by an arbitrary waveform generator with a 1 \SI{}{\giga Sa/s} sampling rate and mixed with a continuous microwave source. Simultaneously, a continuous probe tone set to the readout resonator frequency is used to dispersively measure the qubit state. The outgoing signal is then demodulated and a homodyne detection voltage $V_H$ is measured by a digitizer with a 500 \SI{}{\mega Sa/s} sampling rate. 

\section*{Appendix D: Aluminum surface morphology}
The aluminum film morphology depends on the thickness of the film which leads to noticeable changes in transport properties. In Fig. 9 we show Atomic force microscopy images of samples with varying thicknesses of aluminum. The images are 5$\times$\SI{5}{\micro m}. The Aluminum grains can clearly be seen. We find that the thinnest sample of 30 nm thickness has the smallest Al grain sizes, and the grain sizes increase as the thickness of the film increases. This is consistent with the rf measurements of resonators on these samples: as the film thickness decreases and the Al grain size decreases, the mean free path in the superconductor $l$ decreases, increasing the London penetration depth in the dirty limit where the coherence length $\xi_0\gg l$. This leads to an increase in the kinectic inductance and in turn an increase in inductive loss. 

\bibliography{references_shabani_growth}

\end{document}